\documentclass[pra,aps,showpacs,twocolumn,superscriptaddress]{revtex4-1}
\usepackage{amsmath}
\usepackage{amssymb}
\usepackage{graphicx}
\usepackage{epstopdf}
\usepackage{easybmat}
\usepackage{color,soul}
\usepackage{times,txfonts}
\usepackage{ulem}
\usepackage{mathtools}
\newcommand{\ket}[1]{|#1\rangle}
\newcommand{\bra}[1]{\langle #1|}
\usepackage[bookmarks=false]{hyperref}
\hypersetup{colorlinks=true,citecolor=blue,linkcolor=blue,urlcolor=blue,pdfstartview=FitH,bookmarksopen=true}


\begin{document}

\title {Dynamical decoupling protection for three-level systems}
\author{P. Z. Zhao}
\email{pzzhao@sdu.edu.cn}
\affiliation{Department of Physics, Shandong University, Jinan 250100, China}
\author{Lei Qiao}
\email{qiaolei@buaa.edu.cn}
\affiliation{Peng Huanwu Collaborative Center for Research and Education, Beihang University, Beijing 100191, China}
\date{\today}

\begin{abstract}
In addition to the traditional two-level system, the three-level system serves as another important elemental building block for the manipulation of qubits. However, the quantum information processing in the three-level system is also subject to the decoherence induced by the interaction between the quantum system and its environment or by the crosstalk between different qutrits. In this work, we construct a sequence of physically feasible dynamical decoupling operators for the three-level system to mitigate not only the transverse dephasing between the excited state and ground states but also the  longitudinal relaxation among them. Combining the Hamiltonian engineering and our constructed dynamical decoupling sequence, we further realize the dynamical decoupling protection of qutrit-based quantum gates. Our scheme can effectively enhance the fidelity of three-level-based quantum gates through filtering out the environmental noises, which may provide a new horizon to improve the accuracy of three-level-based quantum manipulation.
\end{abstract}

\maketitle

\section{Introduction}

A qubit is physically represented by a two-level system, but driving a two-level system directly might be challenging due to some inaccessible factors. For example, the single-photon transition between two ground-qubit states in different hyperfine levels or in the same manifold but with different Zeeman sublevels is forbidden for electric dipole radiation.
The three-level system provides an alternative solution to this problem, where two ground qubit states are coupled indirectly through an excited state. Indeed, the three-level system admits an additional dimension compared with the two-level system, thereby promising a potential superiority for storing and manipulating quantum information. For this reason, the three-level system has become an important elemental building block for the manipulation of qubits.
In many physical systems, the three-level system has been utilized to realize various quantum information processing tasks, such as coherent population transfer \cite{Bergmann,Vitanov}, entanglement distribution \cite{Cirac} and gate operations \cite{Duan,Sjoqvist,Xu}.
Especially in the neutral-atom system, the high-lying Rydberg state combined with two ground states consists of a three-level $\Lambda$ configuration, where fast entangling gates can be realized by coupling the Rydberg levels of two qutrits using the Rydberg blockade or Rydberg artiblockade \cite{Jaksch,Luhin,Ates,Saffman,Shao}.
Nevertheless, the quantum information processing in the three-level system is still subject to the  decoherence induced by the interaction between the quantum system and its environment
or by the crosstalk between different qutrits.

Dynamical decoupling (DD) provides an active method to mitigate the decoherence, which arises from the protection of two-level systems \cite{Viola,Viola1999}.
It operates through averaging out the undesired system-environment interaction by applying a sequence of fast and strong symmetrizing pulses to the qubit system.
Hence, one of the key aspects for the DD protection is to identify and construct a physically realizable DD group that targets specific types of environmental noises.
DD can be used to protect quantum memories and quantum gates. For the protection of quantum memories, the periodic DD (PDD) \cite{Viola2003} eliminates the system-environment interaction to the first order (e.g., in the sense of a Magnus expansion) while concatenated DD (CDD) \cite{Khodjasteh} and Uhrig DD (UDD) \cite{Uhrig,Yang,West,Wang} can offer higher-order elimination.
CDD is implemented by utilizing the PDD as a basic building block and then recursively concatenating the PDD sequence. Consequently, the number of consuming pulses increases exponentially as the decoupling order increases. UDD reduces the expenditure through replacing the equidistant sequence by nonequidistant sequence so as to realize the higher-order decoupling within the linear overhead of consuming pulses. Hence, UDD represents a significant advancement in the field of decoherence suppression.

However, the aforementioned routes cannot be directly translated to the protection of quantum gates. This is because the protection of quantum memories focus mainly on the static qubit and hence does not include the time evolution driven by the external field. But for the protection of quantum gate, the dynamical process is driven by the driving Hamiltonian. If the decoupling operators are artificially inserted into dynamical process of the quantum system, they will interfere with the gate operations and hence alter the native time evolution of the quantum system.
To resolve this problem, the logical-qubit-based approaches were proposed but their physical realization related to particular dynamical symmetries remains experimentally challenging \cite{Zanardi,Viola2000,Lidar,West2010}.
Another approach towards DD protection of quantum gates resorts to the average Hamiltonian theory with inserting the gate operations between equidistant DD pulses, but only providing the first-order protection
\cite{Souza,Zhang2014,Suter2016}. A recent approach towards the higher-order protection of quantum gates was put forward by combining Hamiltonian engineering and DD sequences for the quantum memory protection such as CDD and UDD \cite{Gong}.
This approach makes the DD protection of quantum memories and quantum gates unified and applied
on an equal footing. In addition to theoretical progress, experimental demonstrations of DD protection for quantum systems have been also highly successful \cite{Biercuk,Du,Ryan,deLange,Wang2011,Alvarez,Ezzell}.

However, the aforementioned advances focused mainly on the two-level system and have not yet been well established in the three-level system.
Considering the complicity of the three-level system, engineering the DD protection of qutrit memories and qutrit gates is not trivial at all.
It is interesting to note that some DD sequences for the three-level system have already been proposed based on the Heisenberg-Weyl group \cite{Vitanov2015,Du2022,ADlidan,Iiyama}, but cycling the control propagator through all the elements of the Heisenberg-Weyl group is very ineffective in the practical realization. Moreover, these schemes primarily target the cancellation of phase diffusion.
It is therefore highly desirable to develop the DD theory in the three-level system regarding to a relatively general environment with a simple quantum control.
In this work, we construct a sequence of physically feasible DD operators for the three-level system to mitigate not only the transverse dephasing between an excited state and two ground states but also the longitudinal relaxation arising from the transverse noise among the excited state and ground states.
Combining the Hamiltonian engineering and our constructed DD operators, we further realize the DD protection of three-level-based quantum gates. Computational demonstration indicates that our DD scheme can effectively filter out the leading order of environmental noises in the three-level system so that the fidelity of qutrit gates can be obviously enhanced. Our work provides an additional horizon to improve the accuracy of three-level-based quantum manipulation.

\section{Model}

Consider a three-level quantum system consisting of two ground states $\ket{0}$ and $\ket{1}$ and an excited state $\ket{e}$, which form the $\Lambda$ configuration.
The two ground states are generally taken as different hyperfine levels or different Zeeman sublevels in the same manifold of an atom or ion. This implies that the coupling between the two ground qubit states is forbidden for electric dipole radiation.
The quantum system is coupled to its environment with the total Hamiltonian
\begin{align}\label{eq1}
H_{\mathrm{tot}}(t)=H_{0}(t)+H_{\mathrm{E}}+H_{\mathrm{I}},
\end{align}
where $H_{0}(t)$ is the system Hamiltonian, $H_{\mathrm{E}}$ is the environment Hamiltonian and $H_{\mathrm{I}}$ is the interaction Hamiltonian describing the coupling between the system and its environment.
Similar to the phase and amplitude errors occurring in the two-level qubit system, it is reasonable to assume that the directly coupled levels for the three-level system also include a transverse dephasing error
\begin{align}\label{eq2}
H_{\mathrm{IP}}=(\ket{e}\bra{e}-\ket{0}\bra{0})\otimes{B_{0}}
+(\ket{e}\bra{e}-\ket{1}\bra{1})\otimes{B_{1}}
\end{align}
and a longitudinal relaxation error,
\begin{align}\label{eq3}
H_{\mathrm{IL}}=&\left(e^{-i\phi_{0}}\ket{0}\bra{e}+e^{i\phi_{0}}\ket{e}\bra{0}\right)\otimes{E_{0}}
\notag\\
&+\left(e^{-i\phi_{1}}\ket{1}\bra{e}+e^{i\phi_{1}}\ket{e}\bra{1}\right)\otimes{E_{1}}.
\end{align}
Here, the left terms of the tensor product represents the system operators while the corresponding $B_{0}$, $B_{1}$, $E_{0}$ and $E_{1}$ denote the related environmental operators.
It is worth emphasizing that we ignored the the system-environment interaction including the coupling between the ground states. This is because the two ground states are more stationary than the excited state so that the decoherence between the ground states is much smaller than the one between the excited state and the ground states. For example, in trapped ions, the coherence time of the hyperfine qubit  exceeded $10$ minutes \cite{Wang2017} and the fidelity of single-qubit operations facilitated by the microwave manipulation has reached the fault-tolerant threshold of quantum error correction \cite{Harty}.

\section{Decoupling operations}

Let us first present the DD protection of quantum memories in three-level system without accounting for the system Hamiltonian. To average out the system-environment interaction, we need to apply an external control field $H_{\mathrm{C}}(t)$ to the quantum system. In this case, the total Hamiltonian can be rewritten as $H_{\mathrm{tot}}(t)=H_{\mathrm{C}}(t)+H_{\mathrm{E}}+H_{\mathrm{I}}$. In the framework with respect to the applied external field $H_{\mathrm{C}}(t)$, the effective total Hamiltonian is given by $H^{\mathrm{eff}}_{\mathrm{tot}}(t)=U^{\dagger}_{\mathrm{C}}(t)H_{\mathrm{I}}(t)U_{\mathrm{C}}(t)+H_{\mathrm{E}}$, where the unitary operator $U_{\mathrm{C}}(t)\equiv\mathcal{T}\exp[-i\int^{t}_{0}H_{\mathrm{C}}(t^{\prime})\mathrm{d}t^{\prime}]$ is completely determined by the external field $H_{\mathrm{C}}(t)$. The evolution operator then yields $U_{\mathrm{tot}}(t)=U_{\mathrm{C}}(t)U^{\mathrm{eff}}_{\mathrm{tot}}(t)$, where $U^{\mathrm{eff}}_{\mathrm{tot}}(t)\equiv\mathcal{T}\exp[-i\int^{t}_{0}H^{\mathrm{eff}}_{\mathrm{tot}}(t^{\prime})\mathrm{d}t^{\prime}]$ represents the effect from the system-environment interaction.
If the external filed is chosen to satisfy the condition (i) $U_{\mathrm{C}}(t+\tau)=U_{\mathrm{C}}(t)$ with the period of time $\tau$, the evolution operator is reduced to $U_{\mathrm{tot}}(t)=U^{\mathrm{eff}}_{\mathrm{tot}}(t)$. If the external field is further taken to satisfy the condition (ii) $\mathcal{T}\exp[-i\int^{\tau}_{0}U^{\dagger}_{\mathrm{C}}(t)H_{\mathrm{I}}(t)U_{\mathrm{C}}(t)\mathrm{d}t]=I$, the first-order effect induced by the system-environment interaction shall be eliminated from $U^{\mathrm{eff}}_{\mathrm{tot}}(t)$ and hence the evolution operator yields $U_{\mathrm{tot}}(t)=\exp(-iH_{\mathrm{E}}\tau)$. This implies that up to the leading order, the environment is completely decoupled from the quantum system.

The above discussion indicates that to realize DD protection of quantum memories, we need to find an external control filed that satisfies the conditions (i) and (ii). An approach to constructing such an external field is to choose a set of unitary operators $\{p_{k}\}^{N}_{k=1}$ fulfilling $\sum^{N}_{k=1}p^{\dagger}_{k}H_{\mathrm{I}}p_{k}=0$. This operates by using $p^{\dagger}_{k}$ and $p_{k}$ as the propagators and then applying them at the beginning and the end of the time evolution. As such, it is clear that after $N$ evolution periods $N\tau$, the relations $p^{\dagger}_{k}p_{k}=I$ and $\exp(-i\sum^{N}_{k=1}p^{\dagger}_{k}H_{\mathrm{I}}p_{k}\tau)=I$ are fulfilled, i.e., the conditions (i) and (ii) are satisfied. Hence, the set $\{p_{k}\}^{N}_{k=1}$ is a legitimate external field.
Thereafter, a significant problem is how to construct a set of DD operators aiming at the system-environment interaction in Eqs.~(\ref{eq2}) and (\ref{eq3}). For the interaction in Eq.~(\ref{eq2}), it is shown that the system part has the Pauli $Z$-like form.
This motivates us to resort to the Pauli $X$-like decoupling operators, which might be related to $\ket{e}\bra{0}+\ket{0}\bra{e}$ and $\ket{e}\bra{1}+\ket{1}\bra{e}$. However, as indicated below, it is not enough to use the two operators with such a simple construction. For the interaction in Eq.~(\ref{eq3}), the system part includes the linear superposition of the Pauli $X$-like and $Y$-like terms. This motives us to resort to the Pauli $Z$-like decoupling operators, which might be related to $\ket{e}\bra{e}-\ket{0}\bra{0}$ and $\ket{e}\bra{e}-\ket{1}\bra{1}$. Similarly, such a simple construction cannot operate either. Taking into account all the above considerations, we finally redesign the trial solution of DD operators as
\begin{align}\label{eq4}
p_{1}&=\ket{e}\bra{e}+\ket{0}\bra{1}+\ket{1}\bra{0},
\notag\\
p_{2}&=\ket{1}\bra{1}+\ket{e}\bra{0}+\ket{0}\bra{e},
\notag\\
p_{3}&=\ket{0}\bra{0}+\ket{e}\bra{1}+\ket{1}\bra{e},
\notag\\
p_{4}&=\ket{e}\bra{e}-\ket{0}\bra{0}-\ket{1}\bra{1}.
\end{align}
Clearly, these operators are the unitary operator, therefore they can be used as the propagators to generate the time evolution. Moreover, $p_{1}$, $p_{2}$, $p_{3}$ and $p_{4}$ are the Hermitian operators, implying that these DD propagators can work by applying them at the beginning and end of the time evolution without needing their Hermitian conjugate terms. In the following, we illustrate these DD operators can be used to decouple the quantum system (the quantum memory) from its environment.

First, we show that the DD operators $p_{1}$, $p_{2}$ and $p_{3}$ can be used to eliminate transverse dephasing induced by the interaction $H_{\mathrm{IP}}$.
This is done by applying $p_{1}$, $p_{2}$ and $p_{3}$ to the quantum system in such a way that
$[p_{3}f_{\mathrm{IP}}p_{3}][p_{2}f_{\mathrm{IP}}p_{2}][p_{1}f_{\mathrm{IP}}p_{1}]$, where
$f_{\mathrm{IP}}\equiv\exp(-iH_{\mathrm{IP}}\tau)$ denotes the unitary operator related to the phase-flip error. Because $p_{1}H_{\mathrm{IP}}p_{1}+p_{2}H_{\mathrm{IP}}p_{2}+p_{3}H_{\mathrm{IP}}p_{3}=0$, the phase-flip error can be eliminated up to the first order. It is interesting to note that this sequence does not contains an identity operator and hence excludes the term involving  the unitary operator $f_{\mathrm{IP}}$ alone. Second, we show that the DD operator $p_{4}$ can be used to eliminate the longitudinal relaxation induced by the interaction $H_{\mathrm{IL}}$.
This is done by inserting the DD operator $p_{4}$ into the time evolution of the quantum system in such a way that $p_{4}f_{\mathrm{IL}}p_{4}f_{\mathrm{IL}}$, where
$f_{\mathrm{IL}}\equiv\exp(-iH_{\mathrm{IL}}\tau)$ represents the evolution operator related to the  longitudinal relaxation. Because of $p_{4}H_{\mathrm{IL}}p_{4}+H_{\mathrm{IL}}=0$, the first-order decoherence effect can be obviously averaged out. From the above discussions, it is straightforward to conclude that to eliminate both the transverse dephasing and longitudinal relaxation, we can first apply $p_{4}$ to the quantum system and then nest the resulting time evolution into the second layer of DD protection determined by $\{p_{1},p_{2},p_{3}\}$.
The first layer reads
\begin{align}\label{eq5}
p_{4}e^{-iH_{\mathrm{I}}\tau}p_{4}e^{-iH_{\mathrm{I}}\tau}\equiv\mathcal{V},
\end{align}
and the second layer reads
\begin{align}\label{eq6}
(p_{3}\mathcal{V}p_{3})(p_{2}\mathcal{V}p_{2})(p_{1}\mathcal{V}p_{1}).
\end{align}
As such, the leading-order effect caused by the transverse dephasing and longitudinal relaxation for the directly coupled levels in the three-level system can be completely removed.
This is the illustration of our approach towards the first-order protection of quantum memories in three-level systems.

Let us now turn to the second-order protection of the quantum memory in the three-level system using the CDD approach.
The second-order effect of the system-environment interaction in the three-level system is mainly induced by the
the commutator between the interaction terms $H_{\mathrm{IP}}$ and $H_{\mathrm{IL}}$ and the commutators between the environment operator $H_{\mathrm{E}}$ and the interaction terms $H_{\mathrm{IP}}$ and $H_{\mathrm{IL}}$.
The exact expression of the commutator $[H_{\mathrm{IP}},H_{\mathrm{IL}}]$ reads
\begin{align}\label{eq7}
[H_{\mathrm{IP}},H_{\mathrm{IL}}]
=&\left(e^{i\phi_{0}}\ket{e}\bra{0}-e^{-i\phi_{0}}\ket{0}\bra{e}\right)\otimes(B_{0}E_{0}+E_{0}B_{0})
\notag\\
&+\left(e^{i\phi_{1}}\ket{e}\bra{1}-e^{-i\phi_{1}}\ket{1}\bra{e}\right)\otimes(B_{1}E_{1}+B_{1}E_{1})
\notag\\
&+\left(e^{i\phi_{0}}\ket{e}\bra{0}\otimes{B_{1}E_{0}}-e^{-i\phi_{0}}\ket{0}\bra{e}\otimes{E_{0}B_{1}}\right)
\notag\\
&+\left(e^{i\phi_{1}}\ket{e}\bra{1}\otimes{B_{0}E_{1}}-e^{-i\phi_{1}}\ket{1}\bra{e}\otimes{E_{1}B_{0}}\right),
\end{align}
and the expressions of the other two commutators $[H_{\mathrm{E}},H_{\mathrm{IP}}]$ and $[H_{\mathrm{E}},H_{\mathrm{IL}}]$ read
\begin{align}\label{eq8}
[H_{\mathrm{E}},H_{\mathrm{IP}}]=&(\ket{e}\bra{e}-\ket{0}\bra{0})\otimes[H_{\mathrm{E}},B_{0}]
\notag\\
&+(\ket{e}\bra{e}-\ket{1}\bra{1})\otimes[H_{\mathrm{E}},B_{1}],
\notag\\
[H_{\mathrm{E}},H_{\mathrm{IL}}]=&\left(e^{-i\phi_{0}}\ket{0}\bra{e}+e^{i\phi_{0}}\ket{e}\bra{0}\right)\otimes[H_{\mathrm{E}},E_{0}]
\notag\\
&+\left(e^{-i\phi_{1}}\ket{1}\bra{e}+e^{i\phi_{1}}\ket{e}\bra{1}\right)\otimes[H_{\mathrm{E}},E_{1}].
\end{align}
For the commutator $[H_{\mathrm{IP}},H_{\mathrm{IL}}]$, we can eliminate their effect by applying the DD operator $p_{4}$ to the quantum system because $p_{4}[H_{\mathrm{IP}},H_{\mathrm{IL}}]p_{4}+[H_{\mathrm{IP}},H_{\mathrm{IL}}]=0$.
For the commutators $[H_{\mathrm{E}},H_{\mathrm{IP}}]$ and $[H_{\mathrm{E}},H_{\mathrm{IL}}]$, it is clear that they have the similar structure to $H_{\mathrm{IP}}$ and $H_{\mathrm{IL}}$ and hence their effect can be eliminated by using the aforementioned approach.
Considering the first-order elimination of the decoherence altogether, the DD protection of the quantum memories up to the second order is given by
\begin{align}
&[(p_{4}(p_{3}\widetilde{\mathcal{V}}p_{3})(p_{2}\widetilde{\mathcal{V}}p_{2})(p_{1}\widetilde{\mathcal{V}}p_{1})p_{4}]
\notag\\
\times&(p_{3}\widetilde{\mathcal{V}}p_{3})(p_{2}\widetilde{\mathcal{V}}p_{2})(p_{1}\widetilde{\mathcal{V}}p_{1}),
\end{align}
where $\widetilde{\mathcal{V}}\equiv(p_{3}\mathcal{V}p_{3})(p_{2}\mathcal{V}p_{2})(p_{1}\mathcal{V}p_{1})$ with $\mathcal{V}$ taken as the form in Eq.~(\ref{eq5}). This is the illustration of our approach towards the second-order protection of quantum memories in three-level systems.

\begin{figure}[htb]
  \includegraphics[scale=0.34]{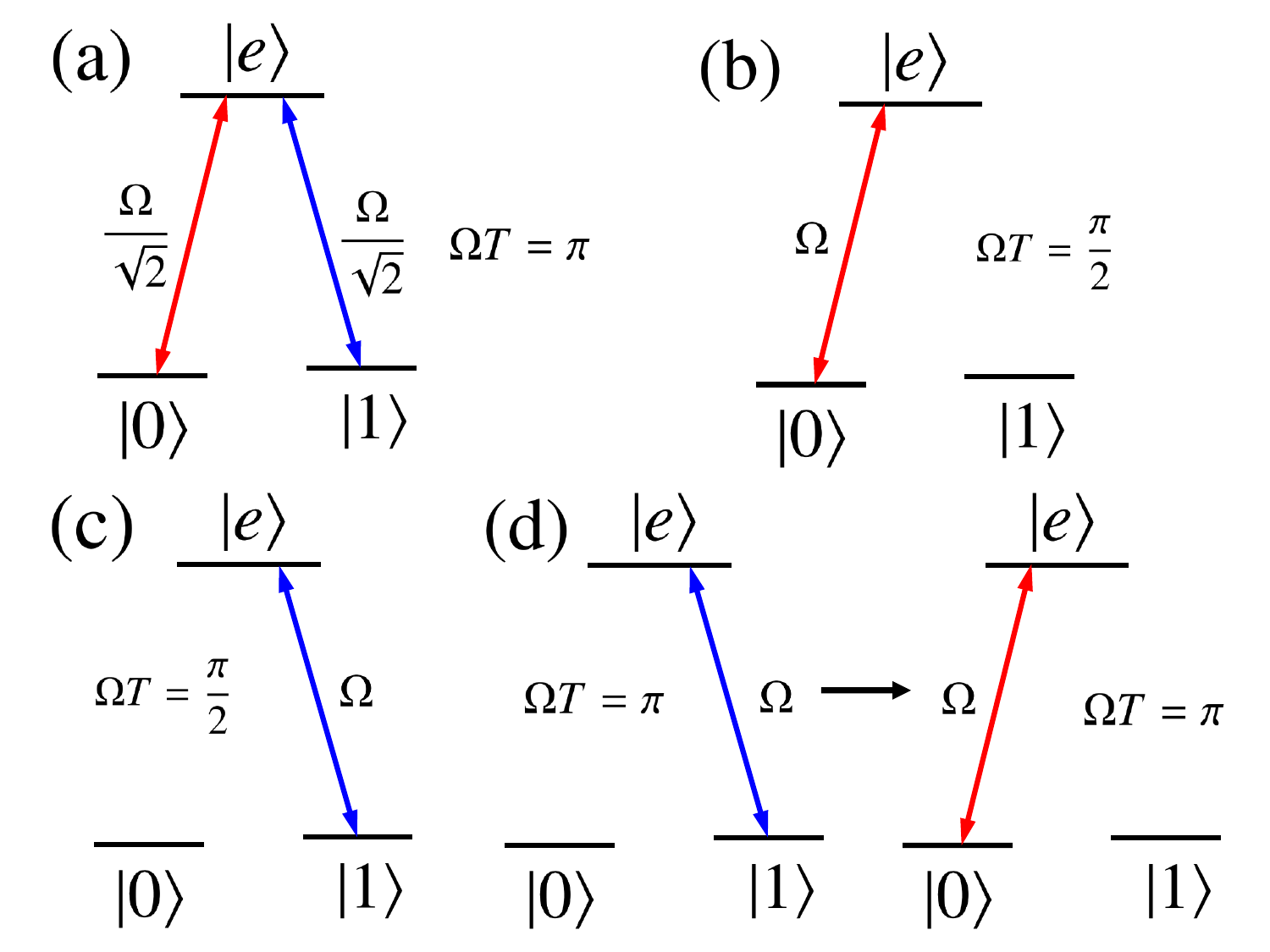}
  \caption{Schematic for the implementation of DD operators $p_{1}$, $p_{2}$, $p_{3}$ and $p_{4}$, depicted by (a), (b), (c) and $(d)$, respectively.}
  \label{Fig1}
\end{figure}

It is worth noting that the DD operators in Eq.~(\ref{eq4}) cannot be simply implemented through a $\pi$ pulse like the qubit case. Hence, after showcasing the DD operators for the protection of three-level systems, we need to then demonstrate how to implement these DD operators in the practical realization.
The DD operator $p_{1}$ can be realized by utilizing the Hamiltonian $\Omega(\ket{e}\bra{0}+\ket{e}\bra{1}+\mathrm{H.c.})/\sqrt{2}$ to drive the quantum system for a period of time $T=\pi/\Omega$, such that $\exp[-i\Omega{T}(\ket{e}\bra{0}+\ket{e}\bra{1}+\mathrm{H.c.})/\sqrt{2}]
\equiv{p_{1}}$, where $\Omega$ is the Rabi frequency. Here, we ignored an unimportant global phase. The schematic for implementation of $p_{1}$ is shown in Fig.~\ref{Fig1}(a).
The DD operator $p_{2}$ can be realized by first applying a $\pi$ pulse to the transition $\ket{0}\leftrightarrow\ket{e}$ with the driving Hamiltonian $\Omega(\ket{e}\bra{0}+\ket{0}\bra{e})$ and then introducing an additional phase to the term $\ket{1}\bra{1}$. The additional phase aims to compensate the phase in the terms $\ket{e}\bra{0}$ and $\ket{0}\bra{e}$, which transfers the relative phase arising from driving the transition $\ket{0}\leftrightarrow\ket{e}$ into a global phase. The DD operator $p_{3}$ can be obtained by using a similar approach as the implementation of $p_{2}$ since the only difference is to exchange the encoding between $\ket{0}$ and $\ket{1}$. The schematic for implementation of $p_{2}$ and $p_{3}$ is shown in Fig.~\ref{Fig1}(b) and (c), respectively.
The DD operator $p_{4}$ can be achieved by using two $2\pi$ pulses to sequently drive the transitions $\ket{0}\leftrightarrow\ket{e}$ and $\ket{1}\leftrightarrow\ket{e}$ such that $\exp[-i\pi(\ket{e}\bra{0}+\ket{e}\bra{0})]\exp[-i\pi(\ket{e}\bra{1}+\ket{e}\bra{1})]
\equiv{p_{4}}$, shown in Fig.~\ref{Fig1}(d). Therefore, as demonstrated above all the DD operators $\{p_{1},p_{2},p_{3},p_{4}\}$ can be realized in the practical implementation.

\section{Protection of quantum gates}

In the above section, we demonstrated the elimination of the system-environment interaction without considering the driving Hamiltonian of the quantum system, that is, we focused mainly on the protection of quantum memories from the environment-induced decoherence. Let us now illustrate how to use the DD sequence to protect the three-level-based quantum gate by taking into account the driving Hamiltonian. In the following, we take the DD protection of the controlled-phases gate as an example.

Consider two three-level systems governed by the driving Hamiltonian $H(t)$.
Each system couples to its environment with the interaction Hamiltonian $H_{\mathrm{I}}$ which includes the terms in Eqs.~(\ref{eq2}) and (\ref{eq3}). To mitigate the docoherence induced by these terms, we apply the DD sequence
$\{p^{\otimes2}_{1},p^{\otimes2}_{2},p^{\otimes2}_{3},p^{\otimes2}_{4}\}$ to the time evolution of the quantum systems. The resulting unitary operator of the total evolution is given by
\begin{align}
\mathcal{U}_{1}=&\prod^{3}_{k=1}
p^{\otimes2}_{k}\left[p^{\otimes2}_{4}\mathcal{T}e^{-i\int^{(k+1)\tau}_{k\tau}H_{\mathrm{tot}}(t)dt}
p^{\otimes2}_{4}\mathcal{T}e^{-i\int^{k\tau}_{(k-1)\tau}H_{\mathrm{tot}}(t)dt}\right]p^{\otimes2}_{k}
\notag\\
=&\prod^{3}_{k=1}\mathcal{T}e^{-i\int^{(k+1)\tau}_{k\tau}(p_{k}p_{4})^{\otimes2}H(t)
(p_{4}p_{k})^{\otimes2}dt}
\mathcal{T}e^{-i\int^{k\tau}_{(k-1)\tau}p_{k}^{\otimes2}H(t)p_{k}^{\otimes2}dt}
\notag\\
&\times e^{-i\tau\sum^{3}_{l=1}p_{l}^{\otimes2}(p^{\otimes2}_{4}H_{\mathrm{I}}p^{\otimes2}_{4}
+H_{\mathrm{I}})p_{l}^{\otimes2}-i6H_{\mathrm{E}}\tau }+\mathcal{O}(\tau^2),
\end{align}
where the product with the smallest $k$ is placed at the right most and $\mathcal{T}$ denotes the time ordering.
From the aforementioned discussion, we can easily obtain the relation $\sum^{3}_{k=1}p_{k}^{\otimes2}(p^{\otimes2}_{4}H_{\mathrm{I}}p^{\otimes2}_{4}+H_{\mathrm{I}})
p_{k}^{\otimes2}=0$. It follows that the exponential term of the system part in the last line disappears and hence we have
\begin{align}
\mathcal{U}_{1}
=&\prod^{3}_{k=1}\mathcal{T}e^{-i\int^{(k+1)\tau}_{k\tau}(p_{k}p_{4})^{\otimes2}H(t)
(p_{4}p_{k})^{\otimes2}dt}
\mathcal{T}e^{-i\int^{k\tau}_{(k-1)\tau}p_{k}^{\otimes2}H(t)p_{k}^{\otimes2}dt}
\notag\\
&\otimes e^{-i6H_{\mathrm{E}}\tau }+\mathcal{O}(\tau^2).
\end{align}
Clearly then, the leading-order effect of system-environment interaction is eliminated. However, the time evolution governed by the driving Hamiltonian $H(t)$ is altered by the applied DD operators. Indeed, the DD operators update the time evolution through changing the quantum driving in such a way that
$H(t)\rightarrow p_{k}^{\otimes2}H(t)p_{k}^{\otimes2}$ and
$H(t)\rightarrow(p_{k}p_{4})^{\otimes2}H(t)(p_{4}p_{k})^{\otimes2}$. The altered effective Hamiltonian is quite different from the original one.

To counteract the unwanted interference from the active DD pulses while still maintaining the DD protection of the controlled-phase gate, we need to engineer the driving Hamiltonian of the quantum system.
To this end, we denote the effective Hamiltonian updated by DD operators in different time intervals as
$\mathcal{H}^{1}_{\mathrm{eff}}\equiv p^{\otimes2}_{1}H(t)p^{\otimes2}_{1}$,
$\mathcal{H}^{2}_{\mathrm{eff}}\equiv(p_{1}p_{4})^{\otimes2}H(t)(p_{4}p_{1})^{\otimes2}$,
$\mathcal{H}^{3}_{\mathrm{eff}}\equiv p^{\otimes2}_{2}H(t)p^{\otimes2}_{2}$,
$\mathcal{H}^{4}_{\mathrm{eff}}\equiv(p_{2}p_{4})^{\otimes2}H(t)(p_{4}p_{2})^{\otimes2}$,
$\mathcal{H}^{5}_{\mathrm{eff}}\equiv p^{\otimes2}_{3}H(t)p^{\otimes2}_{3}$
and $\mathcal{H}^{6}_{\mathrm{eff}}\equiv(p_{3}p_{4})^{\otimes2}H(t)(p_{4}p_{3})^{\otimes2}$.
Then, we engineer the driving Hamiltonian $H(t)$ corresponding to the effective Hamiltonian as
$H_{1}(t)=H_{2}(t)=\Omega(t)\ket{ee}\bra{00}+\mathrm{H.c.}$,
$H_{5}(t)=H_{6}(t)=\Omega(t)\ket{11}\bra{ee}+\mathrm{H.c.}$, and $H_{3}(t)=H_{4}(t)=0$.
In such a way of engineering, the effective Hamiltonians $\mathcal{H}^{1}_{\mathrm{eff}}$, $\mathcal{H}^{2}_{\mathrm{eff}}$, $\mathcal{H}^{5}_{\mathrm{eff}}$ and $\mathcal{H}^{6}_{\mathrm{eff}}$ in the interval $t\in[0,2\tau]\cup[4\tau,6\tau]$ become the same one such that
$\mathcal{H}^{1}_{\mathrm{eff}}=\mathcal{H}^{2}_{\mathrm{eff}}=\mathcal{H}^{5}_{\mathrm{eff}}
=\mathcal{H}^{6}_{\mathrm{eff}}=\Omega(t)\ket{ee}\bra{11}
+\mathrm{H.c.}$, and the effective Hamiltonians in the interval $t\in(2\tau,4\tau)$ yields $\mathcal{H}^{3}_{\mathrm{eff}}=\mathcal{H}^{4}_{\mathrm{eff}}=0$. If we further require
$\int^{6\tau}_{0}\Omega(t)dt=\pi$, the computational state $\ket{11}$ will undergo a cyclic evolution and acquire a $\pi$ phase. Consequently, the first-order protection of the controlled-phase gate is realized. It reads
\begin{align}\label{eq9}
U=\ket{00}\bra{00}+\ket{01}\ket{01}+\ket{10}\bra{10}-\ket{11}\bra{11}.
\end{align}
This completes our demonstration of DD protection of the controlled-phase gate.
\begin{figure*}[htb]
  \includegraphics[scale=0.35]{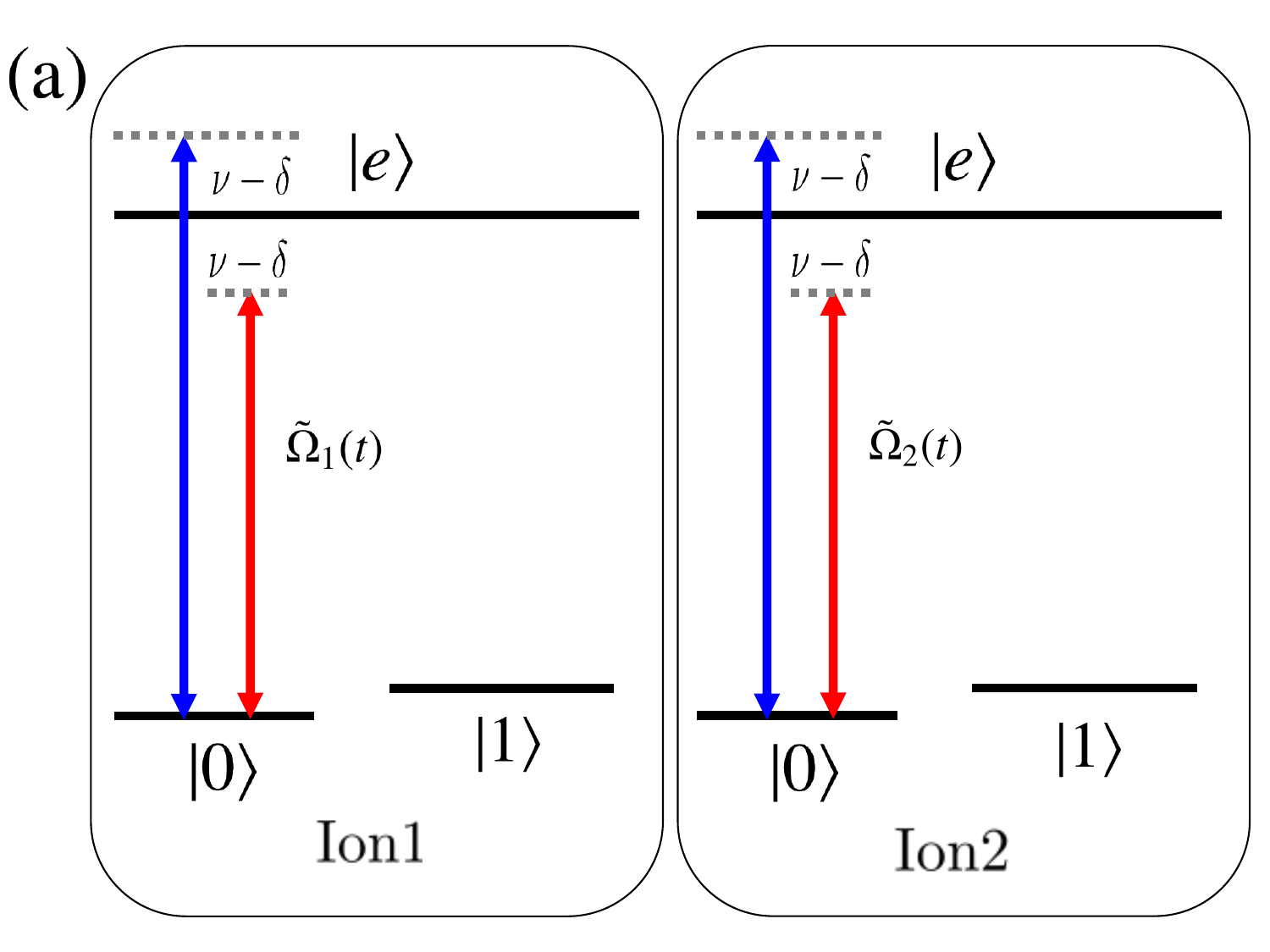}
  \includegraphics[scale=0.35]{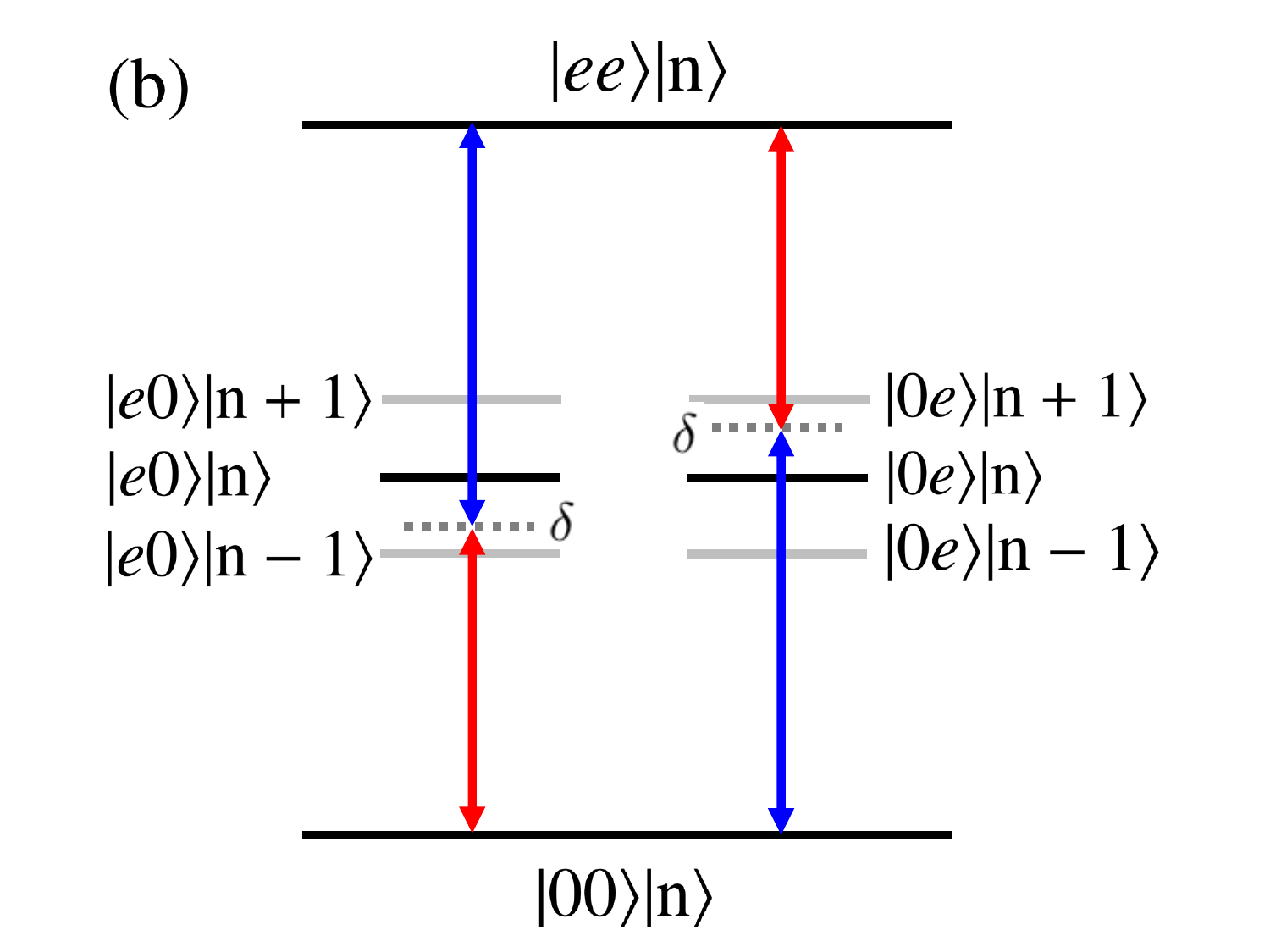}
  \caption{Schematic for the physical implementation of the engineered Hamiltonian. (a) A pair of red and blue sideband lasers with detunings $\nu-\delta$ and $-(\nu-\delta)$ and a common Rabi frequency $\tilde{\Omega}_{1}(t)$ are applied to the first ion and another pair of red and blue sideband lasers with the detunings $\nu-\delta$ and $-(\nu-\delta)$ and a common Rabi frequency $\tilde{\Omega}_{2}(t)$ are applied to the second ion, both of which are used to drive the transition $\ket{0}\leftrightarrow\ket{e}$. (b) Level diagram for the collective transition of the two ions. The single-ion transition $\ket{0}\leftrightarrow\ket{e}$ is strongly suppressed under the large detuning condition whereas the double-ion transition $\ket{00}\leftrightarrow\ket{ee}$ is facilitated by exchanging the vibrational energy between two ions.}
  \label{Fig2}
\end{figure*}

Finally, we show that the engineered driving Hamiltonian for the DD protection of quantum gate is available in the practical physical systems.
In the above discussions, we can see that $H_{1}(t)=H_{2}(t)$, $H_{5}(t)=H_{6}$, and $H_{1}(t)$ differs from $H_{5}(t)$ by only an encoding swap $\ket{0}\leftrightarrow\ket{1}$ and a parameter change $\Omega(t)\leftrightarrow\Omega^{*}(t)$. This indicates that $H_{5}(t)$ can be implemented with a similar approach to $H_{1}(t)$.
Additionally, $H_{3}(t)=H_{4}(t)=0$ represents a delay but not a driving field. Therefore, we only need to demonstrate how to realize the driving Hamiltonian $H_{1}(t)$.
In trapped ions, this Hamiltonian can be achieved easily by using bichromatic lasers with the renowned M{\o}lmer-S{\o}rensen coupling \cite{MS,SM,SM2000}.
Specifically, we apply a pair of red and blue sideband lasers with detunings $\nu-\delta$ and $-(\nu-\delta)$ and a common Rabi frequency $\tilde{\Omega}_{1}(t)$ to the first ion and another pair of  red and blue sideband lasers with detunings $\nu-\delta$ and $-(\nu-\delta)$ and a common Rabi frequency $\tilde{\Omega}_{2}(t)$ to the second ion, both of which facilitate the transition $\ket{0}\leftrightarrow\ket{e}$, as is shown in Fig.~\ref{Fig2}(a).
Here, $\nu$ is the frequency of the vibrational mode and $\delta$ is an additional detuning satisfying $\delta\ll\nu$. In the rotating frame and rotating-wave approximation, the Hamiltonian in the Lamb-Dicke regime is given by
$H(t)=i\eta\tilde{\Omega}_{1}(t)(e^{-i\delta t}a+e^{i\delta t}a^{\dagger})
\ket{e}_{11}\bra{0}+i\eta\tilde{\Omega}_{2}(t)(e^{-i\delta t}a+e^{i\delta t}a^{\dagger})
\ket{e}_{22}\bra{0}+\mathrm{H.c.}$,
where $a$ and $a^{\dagger}$ are the annihilation and creation operators of the vibrational mode and $\eta$ is the Lamb-Dicke parameter satisfying $\eta^{2}(n+1)\ll1$ ($n$ is the vibrational quantum number).
In the large detuning condition $\delta\gg|\eta\tilde{\Omega}_{1}(t)|, |\eta\tilde{\Omega}_{1}(t)|$, the single-ion transition $\ket{0}\leftrightarrow\ket{e}$ can be strongly suppressed whereas the double-ion transition $\ket{00}\leftrightarrow\ket{ee}$ is facilitated by exchanging the vibrational energy between two ions, the effective coupling structure of which is shown in Fig.~\ref{Fig2}(b).
As a consequence, the Hamiltonian is reduced to
$H(t)=\Omega(t)\ket{ee}\bra{00}+\mathrm{H.c.}$, where $\Omega(t)=2\eta^{2}\tilde{\Omega}_{1}(t)\tilde{\Omega}_{2}(t)/\delta$.
Here, we ignored the two-ion coupling term between $\ket{0e}$ and $\ket{e0}$ because it neither belongs to the computational subspace nor couples with the computational subspace
and we have also ignored Stark shift terms that can be easily compensated by applying additional lasers. The Hamiltonian described above is just the driving Hamiltonian $H_{1}(t)$, indicating that the driving Hamiltonian $H_{1}(t)$ along with the other driving Hamiltonians $H_{2}(t)$, $H_{5}(t)$ and $H_{6}(t)$ are all physically realizable in the practical three-level system.

\section{Computational demonstration}

Having showcased the theoretical concept, we now computationally demonstrate the performance of our approach to the DD protection of qutrit gates. In the presence of the system-environment interaction, we
assume that the two-qutrit quantum system is subject to a random error \cite{Wu} with the error Hamiltonian $H_{e}=\sum^{2}_{k=1}[\delta^{k}_{0}(t)(\ket{e}_{k}\bra{e}
-\ket{0}_{k}\bra{0})+\delta^{k}_{1}(t)(\ket{e}_{k}\bra{e}-\ket{1}_{k}\bra{1})]
+[\eta^{k}_{0}(t)(\ket{e}_{k}\bra{0}+\ket{0}_{k}\bra{e})
+\eta^{k}_{1}(t)(\ket{e}_{k}\bra{1}+\ket{1}_{k}\bra{e})]
+[\varepsilon^{k}_{0}(t)(-i\ket{e}_{k}\bra{0}
+i\ket{0}_{k}\bra{e})+\varepsilon^{k}_{1}(t)(-i\ket{e}_{k}\bra{1}+i\ket{1}_{k}\bra{e})]$. Here, the first, second and third square bracket terms correspond to the $Z$-, $X$- and $Y$-type errors for the directly coupled levels defined in analogy to the two-level system, and $\delta^{k}_{0}(t)$, $\delta^{k}_{1}(t)$, $\eta^{k}_{0}(t)$, $\eta^{k}_{1}(t)$, $\varepsilon^{k}_{0}(t)$ and $\varepsilon^{k}_{1}(t)$ are the time-dependent error parameters for the $k$th qutrit. For convenience, we set all the error parameters to be the same one
$\kappa(t)=\epsilon\int^{\infty}_{0}d\omega S(\omega)\cos(\omega{t})/\pi$, where $\epsilon$ is a time-independent amplitude and $S(\omega)=\omega\Theta(\pi/2-\omega)$ is the spectral density with $\Theta(\cdot)$ being the Heaviside step function.
The performance is characterized by the fidelity $F=\bra{\phi}\rho\ket{\phi}$, where $\ket{\phi}$ is the target output state and $\rho$ is the real output state.

\begin{figure*}[htb]
  \includegraphics[scale=0.6]{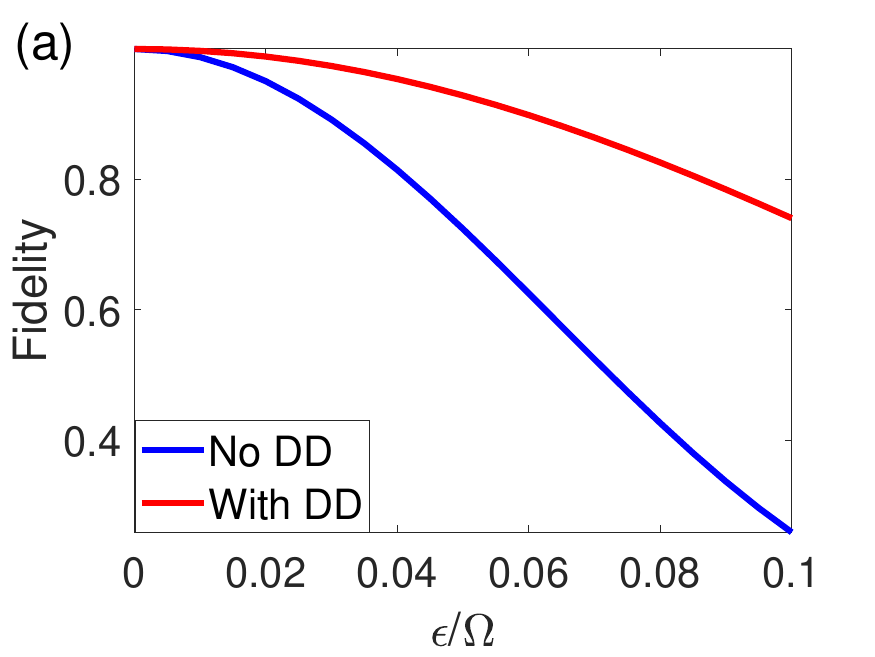}
  \includegraphics[scale=0.6]{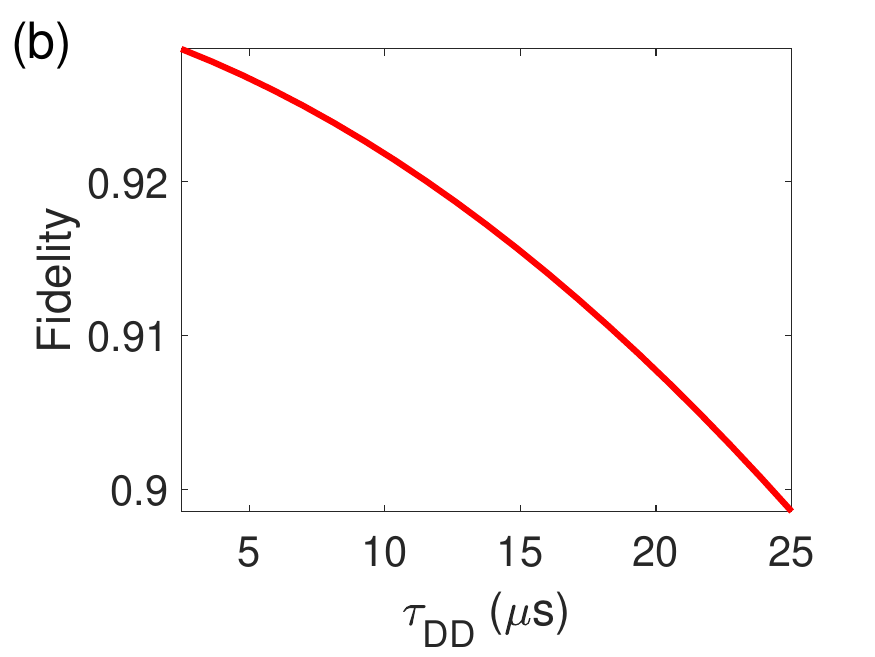}
    \caption{The performance for DD protection of quantum gates using finite-time gate operations as DD pulses. (a) The fidelity of the controlled-phase gate with (red) and without (blue) DD protection decays as the system-environment amplitude error parameter increases over $\epsilon\in[0,0.1\Omega]$. (b) The fidelity of the DD-protected controlled-phase gate versus the time duration of one DD operation, namely, $\tau_{\mathrm{DD}}\in[2.5~\mu{\mathrm{s}},25~\mu{\mathrm{s}}]$, with a fixed $\epsilon=0.05\Omega$.    }
  \label{Fig3}
\end{figure*}

Our goal is to evaluate the performance of the DD protection for the controlled-phase gate depicted in Eq.~(\ref{eq9}) under the influence of decoherence. To this end, we take the input state of the two-qubit system as $(\ket{00}+\ket{11})/\sqrt{2}$ [the corresponding target output state is given by  $\ket{\phi}=(\ket{00}-\ket{11})/\sqrt{2}$] and the amplitude parameter of the coupled two-qubit driving Hamiltonian as $\Omega=2\pi\times1\mathrm{kHz}$. The strength of our square-shaped DD pulse is chosen to be $2\pi\times100\mathrm{kHz}$, indicating that we explore finite-time single-qubit gates rather than the ideal instantaneous pulse operations as our actual DD pulses. The piecewise adjustment of the driving Hamiltonian is executed right after the DD pulses.
To demonstrate the performance of our DD protection scheme in presence of the environmental noise, we further tune the amplitude error parameter $\epsilon$ and then plot the fidelity $F$ of the controlled-phase gate under the DD protection and without any DD protection versus the amplitude error parameter over the range $\epsilon\in[0,0.1\Omega]$, represented by the red and blue lines in Fig.~\ref{Fig3}(a).
The result clearly shows that the quantum gate with DD protection outperforms the bare gate.
Indeed, for a relatively large amplitude strength $\epsilon=0.1\Omega$, i.e., the system-bath coupling strength is about $0.1$ times of the gate Hamiltonian strength, the fidelity of the bare controlled-phase gate is as low as $25.85\%$, indicating the quantum gate is basically destroyed. However, when the DD sequence is applied to the time evolution, the fidelity of quantum gates boosts back to $74.04\%$, implying the effectiveness of our scheme. For a proper strength $\epsilon=0.05\Omega$, the fidelity of the gate with DD protection can be up to $92.86\%$ while the bare gate can only reach $72.38\%$.

In the above numerical simulations, we utilize finite-time gate operations as our DD sequence. Considering that single-qubit operators still need some time to complete its time evolution, the applied DD pulses also experience decoherence arising from the system-environment interaction.
Obviously, if the running time of DD pulses is long, the decoherence effect on the DD operators will be serious and then the effectiveness of DD protection of time evolution will degrade. There is hence a trade-off between the robustness of the DD protection and the duration of the applied DD operations. That is, although ideal DD pulses are considered to convey our physical insights, there is in practice limitation to the DD protection in the duration of the applied pulses. To investigate the influence of DD duration on the performance of DD-protected quantum gates, we plot the fidelity $F$ versus the duration of a single DD operation over $\tau_{\mathrm{DD}}\in[2.5~\mu{\mathrm{s}},25~\mu{\mathrm{s}}]$ (meaning that the coupling strength of DD pulses varies from $10$ to $100$ times of the gate Hamiltonian strength). The result is shown in Fig.~\ref{Fig3}(b) with the fixed system-bath amplitude strength $\epsilon=0.05\Omega$. It follows that as the duration of DD pulse increases from $2.5~\mu{\mathrm{s}}$ to $25~\mu{\mathrm{s}}$, the gate fidelity decays from $92.86\%$ to $89.85\%$. These numerical findings do not represent at all what one can achieve on an actual quantum computing platform, but indicate the promise of our DD protection scheme in enhancing the fidelity of the three-level-based qutrit gates.

\section{Discussion and Conclusion}

The general principle of pulse decoupling has been known already for a long time \cite{Lidar2014}, according to which the PDD sequence containing the Pauli group was proposed to average the system-environment interaction in the qubit system. The DD sequence based on the Heisenberg-Weyl group \cite{Vitanov2015,Du2022,ADlidan,Iiyama} was also developed to suppress the decoherence in the multiple-level system. However, unlike the qubit case, the three-level system has more complex level configuration and hence is subject to more environmental noises. It is therefore more difficult to construct the decoupling operators related to the system-environment interaction in the three-level system. Moreover, although the the group-based approach can eliminate an arbitrary noise but not just a subset of noise, it needs a large number of unitary operators to complete the decoupling task.
Hence, it is also nontrivial to find a DD sequence to reduce the overhead of the pulse number for a given set of noise. In our work, we constructed such a physically feasible DD sequence aiming at the transverse dephasing noise and the longitudinal relaxation noise in the three-level system,  which is not a group but is enough to decouple the quantum system from its environment with only four DD operators. Furthermore, we use the DD sequence to realize the first-order and second-order protections of quantum memories. By combining the Hamiltonian engineering and our constructed DD sequence, we also realized the DD protection of qutrit-based quantum gates. Computational demonstration indicated that our DD scheme can effectively filter out the leading order of environmental noises in three-level systems so that the fidelity of qutrit gates can be conformably enhanced. Note that our scheme is only applicable to the non-Markovian noise rather than the Markovian noise. Further extension of our work may lie in realizing the higer-order protection of qutrit gates using CDD and UDD approaches. This is an important topic for further enhancing the fidelity of qutrit gates but it may be more challenging because the complicity of the three-level system makes the engineering of the accessible DD sequence and the realizable driving Hamiltonian for the higher-order DD protection of qutrit gates more difficult than that of qubit gates based on two-level systems. On the other hand, our DD sequence is based on the square-shaped pulses. If pulse shaping is introduced to the designing of DD pulses, we foresee that our approach to the DD protection might further enhance the performance of qutrit-based quantum gates.

\begin{acknowledgments}
This work was supported by the National Natural Science Foundation of China through Grants No. 12305021.
\end{acknowledgments}

\end{document}